\documentclass[]{aa}
\usepackage{natbib}

\usepackage{graphicx}
\usepackage[a4paper]{hyperref}
\usepackage[a4paper]{hyperref}
\usepackage{psfig}
\begin{document}
\idline{000}{000}
\doi{000}
\def\teff{$T\rm_{eff }$}
\def\kms{${\mathrm {km s^{-1}}}$}

\title{The lithium content of the globular cluster
NGC 6397
\thanks{Based on
observations made with the 
ESO VLT-Kueyen telescope at the Paranal Observatory, Chile,
in the course
of the ESO-Large program 165.L-0263
                            }}

   \subtitle{}

\author{
P. \,Bonifacio \inst{1},
L. Pasquini\inst{2},
F. Spite\inst{3},
A. Bragaglia\inst{4},
E. Carretta\inst{5},
V. Castellani\inst{6},
M. Centuri\`on\inst{1},
A. Chieffi\inst{7},
R. Claudi\inst{5},
G. Clementini\inst{4},
F. D'Antona\inst{8},
S. Desidera\inst{5,9},
P. Fran\c cois\inst{3},
R.G. Gratton\inst{5},
F. Grundahl\inst{10},
G. James\inst{3},
S. Lucatello\inst{5,9},
C. Sneden\inst{11},
\and
O. Straniero\inst{12}
          }
\offprints{P. Bonifacio,\email{bonifaci@ts.astro.it}}

\institute{
Istituto Nazionale di Astrofisica --
Osservatorio Astronomico di Trieste, Via G.B.Tiepolo 11, 
I-34131 Trieste, Italy 
\and
European Southern Observatory, Karl--Schwarzschild--Str. 2, D-85748 
Garching bei M\"unchen, Germany
\and
Observatoire de Paris- Meudon, 61, Av. de l'Observatoire, 75014 Paris, France
\and
Istituto Nazionale di Astrofisica --
Osservatorio Astronomico di Bologna, via Ranzani 1, 40127 Bologna, Italy
\and
Istituto Nazionale di Astrofisica --
Osservatorio Astronomico di Padova, Vicolo dell'Osservatorio 5, 35122
 Padova, Italy
\and
Dipartimento di Fisica, Universit\`a di Pisa, 
Piazza Torricelli 2, 56100, Pisa, Italy
\and
Istituto di Astrofisica Spaziale, CNR, 
via Fosso del Cavaliere, 00133 Roma,
Italy
\and
Istituto Nazionale di Astrofisica --
Osservatorio Astronomico di Roma, Italy
Via Frascati 33, 00040 Monteporzio Catone, Roma,
\and
Dipartimento di Astronomia, Universit\`a di Padova, 
vicolo dell'Osservatorio 2, 35122 Padova, 
Italy
\and
Department of Astronomy, University of Aarhus, 
Bygning 520, DK-8000 Aarhus C, Denmark
\and
Department of Astronomy and McDonald Observatory, 
University of Texas at Austin, RLM 15.308, C-1400, Austin, TX,
78712-1083, USA
\and
Istituto Nazionale di Astrofisica --
Osservatorio Astronomico di Collurania-Teramo,
64100 Teramo, 
Italy
}

\authorrunning{Bonifacio et al.}
\mail{bonifaci@ts.astro.it}

\titlerunning{Lithium in NGC 6397}

\date{Received  / Accepted }

\abstract{
We make use of high resolution, high signal-to-noise ratio
spectra of 12 turn-off stars in the metal-poor globular cluster NGC 6397
to measure its lithium content. 
We conclude  that they all have the same lithium
abundance A(Li)$ = 2.34 $ 
with a standard deviation of 0.056 dex.
We use this result, together with Monte Carlo simulations,
to estimate that the maximum allowed intrinsic scatter
is of the order of 0.035 dex.
This is a new stringent
constraint to be fulfilled by
stellar models  which predict Li depletion.
We argue that although a mild depletion of 0.1 -- 0.2 dex, 
as predicted by recent models,
cannot be ruled out, there is no compelling reason
for it.  This fact, together with the good agreement with
the Li abundance observed in field stars,
supports the primordial origin of lithium in
metal-poor stars. 
Taking the above value as the primordial
lithium abundance implies a cosmic
baryonic density which is either
$\Omega_b h^2 = 0.016 \pm 0.004$
or $\Omega_b h^2 = 0.005 ^{+0.0026}_{-0.0006}$,
from the predictions of standard big bang nucleosynthesis. 
The high baryonic density solution is in agreement
with recent results on the primordial 
abundance of  deuterium and $\rm ^3He$
and on the estimates derived from the fluctuations of the
cosmic microwave background. 
\keywords{
Diffusion -
Stars: abundances -
Stars: atmospheres -
Stars: Population II -
(Galaxy:) globular clusters: NGC 6397 - 
Cosmology: observations }
}
\maketitle{}           

\section{Introduction}

The lithium abundance in metal--poor  turn-off stars 
appears as 
a {\em plateau}, independent of metallicity and effective temperature
\citep{s82a,s82b}, the so-called
{\em Spite plateau}.
Currently its most widely accepted interpretation 
is that the lithium observed in metal--poor stars has been
produced in the big bang and is thus a probe of the universal
baryonic density.
Recent claims that the {\em 
Spite plateau} does not reflect the true primordial abundance 
which would be either {\em lower}, due to chemical evolution effects
\citep{ryan99}, or {\em higher} due to Li depletion
\citep{pin99,pin01,sal01,theado}, deal  with second order effects. The
primordial Li abundance is  within $\sim 
0.2 $ ~dex of the plateau value.
Thus there is a general consensus that the bulk of lithium in
metal--poor stars is of primordial origin.  
If this is the case we expect metal-poor stars, wherever in the universe,
to show the same Li abundance. 
The present instrumentation is not capable of  
providing Li abundances of 
turn-off stars in nearby external galaxies, however
\citet{molaro_casa} suggested 
that extra-galactic Li has already been
detected, at the plateau value, in the
blue-metal-poor star CS 22873-139, possibly accreted from an external
galaxy.
In spite of this, our present knowledge rests
on the study of   Li abundances in different
metal-poor populations of our own Galaxy.
The most studied population is the
metal--poor field halo 
\citep[and references therein]{s82b,bm97,ryan99}. \citet*{mbp97} 
have studied
lithium in metal--poor thick disc candidates and found that they
lie on the {\em Spite plateau};
two more thick disc stars lying on the plateau
were noted by  \citet{romano}.

\citet{mp94} and \citet{pm96} were the first to study
Li in turn-off stars in NGC 6397 and found that 
it is consistent with the plateau.
Recently \citet{thev} published Li abundances for seven turn-off
stars of NGC 6397, based on UVES data similar to our own,
confirming the earlier results. 
Boesgaard 
and collaborators have studied lithium 
in the globular clusters M92 \citep{boe98},
M13 and M71 \citep{boe2000} using the 10m Keck telescope 
and found a sizeable scatter in Li abundances
in all three clusters.
 
The globular cluster NGC 6397 is one of the nearest and best studied
ones. It is one of the main targets of the ESO-Large Program
165.L-0263 which we are conducting with the aim of providing accurate
abundances, reddenings and distances  for globular clusters
with a range of metallicities.
In \citet{g2001} we 
provided the abundances of several elements
and reached the conclusion that NGC 6397 is a very  homogeneous
cluster, from the chemical point of view, down to the main-sequence
stars. It is  also quite 
metal--poor ([Fe/H]$= -2.03 \pm 0.02$).
We deferred the study of Li abundances in turn-off stars
to the present paper because we believe lithium deserves special
care. From an observational point of view the Li I 670.8 nm
resonant doublet, which is the only lithium line observed,  is rather
weak, asymmetric and difficult to measure accurately in faint objects.
From the point
of view of the analysis Li is extremely sensible to the adopted
effective temperature, since it is mostly ionized (Li II $> 99.7$\%) 
in the atmospheres
of solar-type stars and  neutral lithium is only a trace species. 
Combining the data from \citet{thev} with our own, the
total number of turn-off stars of NGC 6397 observed at high resolution
with a high S/N ratio is twelve.
Such a large sample allows us to tackle 
the issue of the 
possible intrinsic dispersion of the  Li content among turn-off 
stars in this cluster.

\begin{figure}[h!]
\resizebox{\hsize}{!}{\includegraphics[clip=true]{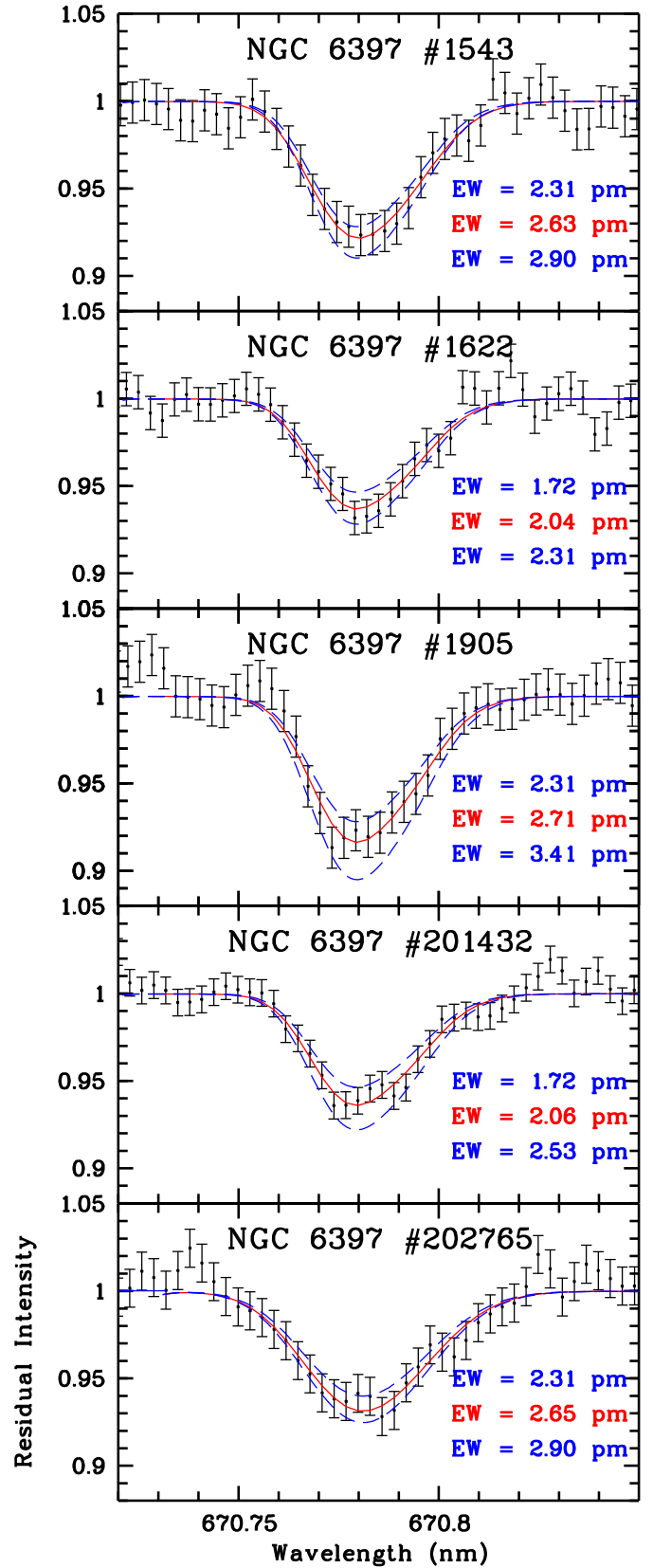}}
\caption{Li doublet for the program stars}
\label{spectra}
\end{figure}

\section{Observations}

Our spectra were collected at ESO-Paranal with 
the UVES spectrograph  \citep{Dekker}
at the Kueyen 8.2m telescope 
between June 15th 2000 and June 22nd 2000, and
have already been described in \citet{g2001}; the log
of the observations is given in Table \ref{log},
the DIMM seeing was noted, when available.
While in \citet{g2001} we have relied on the pipeline,
in this paper we reduced all the spectra beginning
from the raw data using the MIDAS echelle context. 
The purpose was
both to check the accuracy of the pipeline reduction
and to try to improve it, especially trying to get rid
of the low frequency variations in the continuum, at the
level of a few per-cent which were apparent in the pipeline 
data.

The stability of the UVES spectrograph proved to be remarkable:
the flat-fields were perfectly reproducible from one night
to the other, except for the night of the 22nd, for which a slight
shift of the order positions of a couple of pixels appeared;
the wavelength scale was also extremely stable, the largest
peak-to-peak variation in the position of the arc lines observed
on different nights
was 0.005 nm. 
The FWHM of the arc lines in the region of the Li doublet was
0.015 nm, providing a resolution
$R\sim 45000$.

\begin{table}
\caption{Log of the observations}
\label{log}
\begin{tabular}{lcccl}
\hline
\\
star \# & date & UT & $t_{exp}$ & seeing \\
      & d/m/y  & h:m:s & s & arcsec \\
\\
\hline
\\
1543   & 20/06/2000 & 00:56:06 & 4500 & 0.8 \\
1543   & 20/06/2000 & 02:12:44 & 4500 & 0.8 \\
1543   & 20/06/2000 & 03:30:21 & 3600 & 1.0 \\
1543   & 20/06/2000 & 04:33:03 & 4500 & 1.1 \\
1622   & 16/06/2000 & 02:21:00 & 3600 & 0.6 \\
1622   & 16/06/2000 & 03:28:30 & 3600 & 0.55 \\
1622   & 21/06/2000 & 02:50:40 & 3600 & ? \\
1622   & 21/06/2000 & 03:55:12 & 3600 & ? \\
1905   & 19/06/2000 & 00:43:52 & 3600 & 1.0 \\ 
1905   & 19/06/2000 & 01:45:19 & 3600 & 1.0 \\ 
1905   & 22/06/2000 & 00:32:10 & 3600 & 0.8 \\ 
1905   & 22/06/2000 & 01:33:28 & 3600 & 0.8 \\ 
201432 & 18/06/2000 & 05:00:42 & 3600 & 0.8 \\
201432 & 18/06/2000 & 06:05:20 & 3600 & 0.8 \\
201432 & 22/06/2000 & 02:38:09 & 3600 & 0.35 \\
201432 & 22/06/2000 & 03:41:27 & 3600 & 0.35 \\
202765 & 19/06/2000 & 02:50:18 & 3600 & 1.0 \\
202765 & 19/06/2000 & 03:51:39 & 5400 & 1.5 \\
202765 & 21/06/2000 & 00:42:01 & 3600 & 1.1 \\
202765 & 21/06/2000 & 01:43:46 & 3600 & ? \\
\hline
\\
\end{tabular}
\\
{For all the observations the slit was $1''$ and
the CCD binning $2\times2$}
\end{table}

\begin{figure}[t!]
\resizebox{\hsize}{!}{\includegraphics[clip=true]{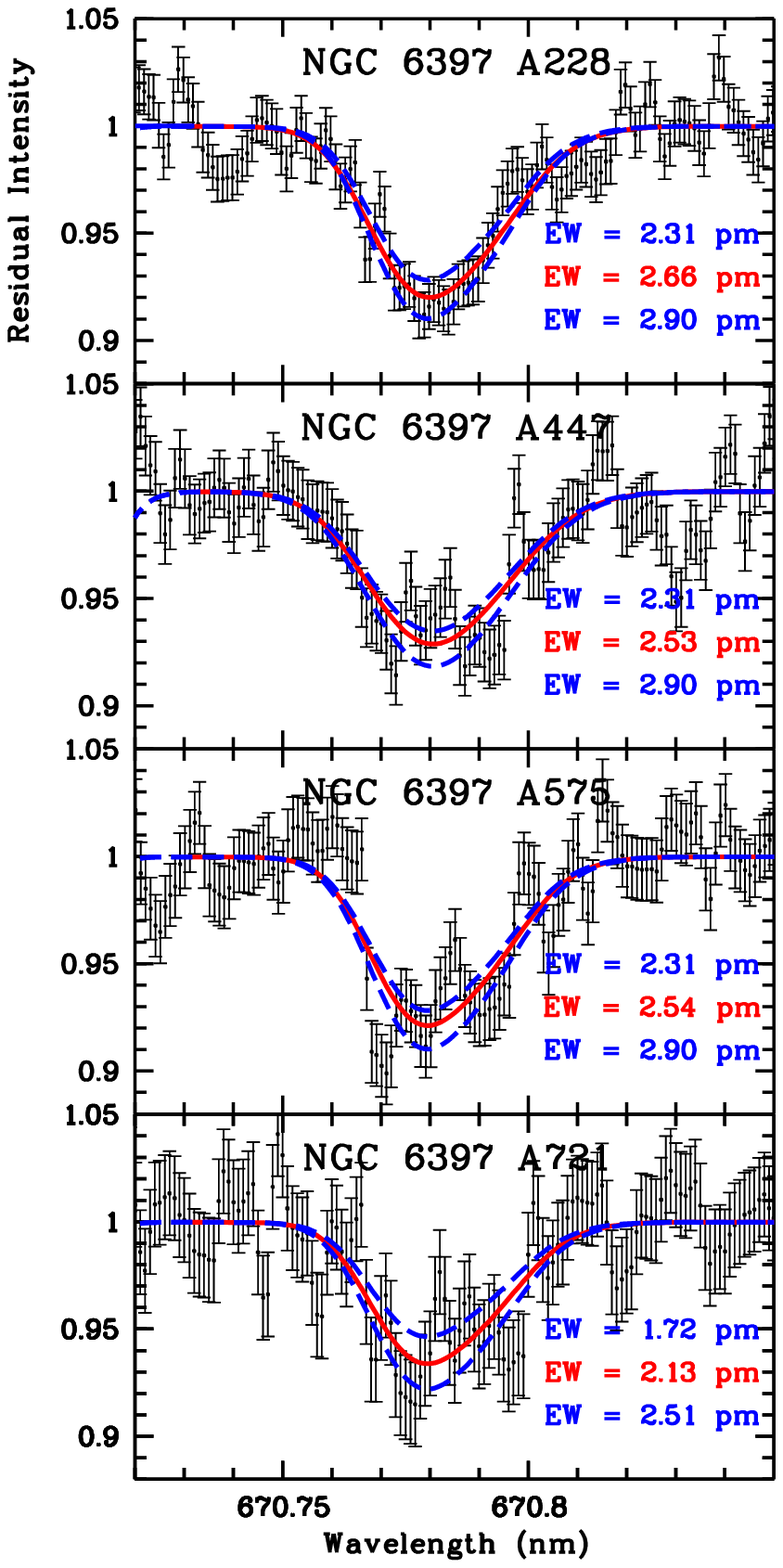}}
\caption{Li doublet for four of the stars of \citet{thev}}
\label{spectra4}
\end{figure}

\begin{center}
\begin{figure}[b!]
\resizebox{\hsize}{!}{\includegraphics[clip=true]{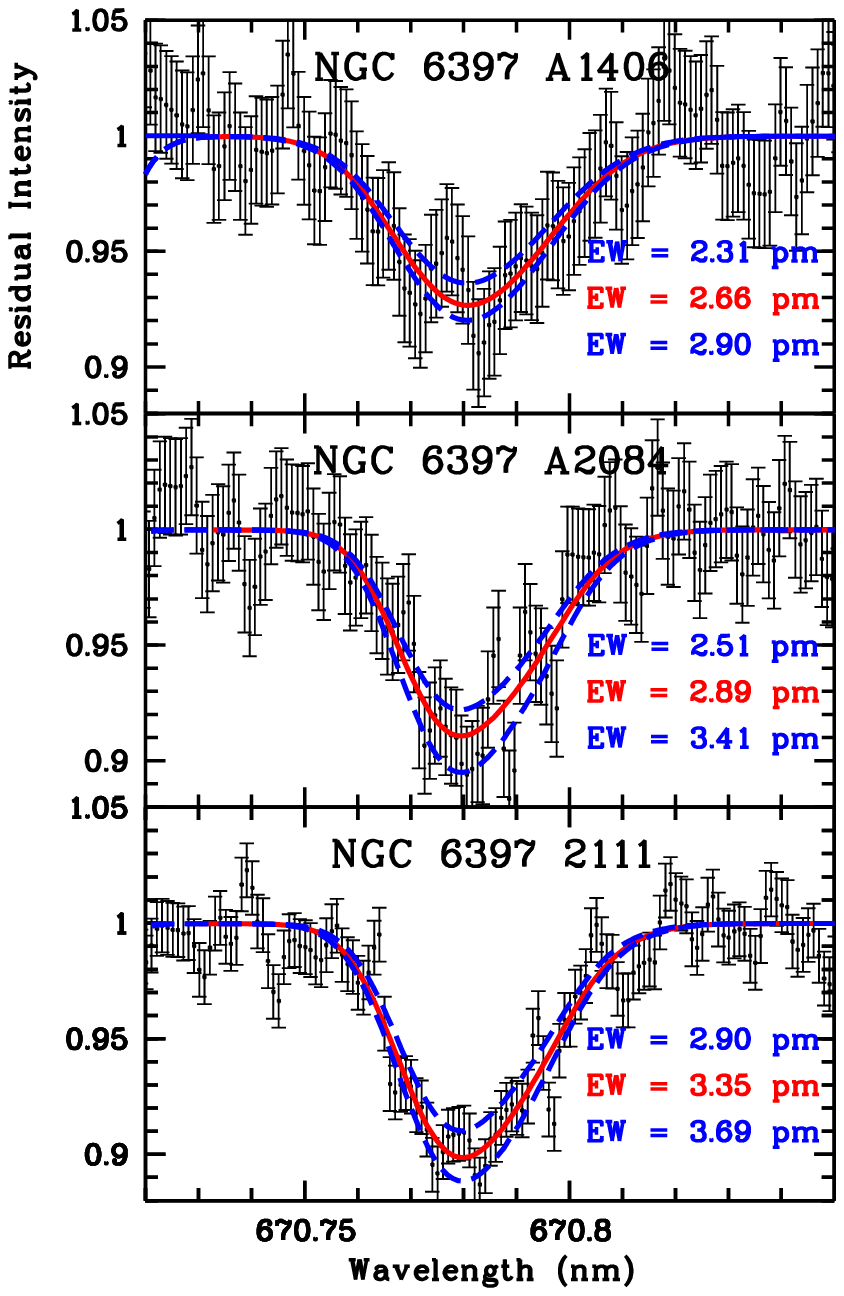}}
\caption{Li doublet for three of the stars of \citet{thev}}
\label{spectra3}
\end{figure}
\end{center}

In spite of our efforts we managed to obtain only a moderate increase
of the S/N ratio, with respect to the pipeline data, showing that
the quality of the pipeline data is very high and perfectly adequate
for scientific analysis. In order  to rectify the spectrum, removing
the signature of the instrument response function, 
we used the flux standard Feige 67, which has been observed on
June the 15th with the same spectrograph setting as the program stars,
except for the slit width which was of $10''$.
The flux data used to calibrate it, was that of \citet{oke} .
The wavelength scale of the Oke data is uncertain by about 0.6 nm,
we therefore arbitrarily shifted his data so that the position
of H$\alpha$ coincided with that in our spectrum. The quality of this
relative flux-calibration 
is of the order of a few percent. 
This was estimated by
comparing the flux of Feige 66, observed on
June 20th and calibrated with the response function derived
from the observations of Feige 67, to the flux of \citet{oke}.
Another spectrum of Feige 66, observed on June 17th is almost identical
to that of June 20th, after a suitable scaling.
This gives us confidence that the instrumental response function is very
stable and reproducible. We stress that the purpose of this calibration
is only to rectify the spectrum and it is meant to be accurate in a
relative, not absolute sense. 
The flux calibration was not entirely successful in removing the 
ondulation in the continuum; in the end we therefore applied a
Fourier high-pass filter to remove it.

\begin{table*}
\caption{Equivalent widths and Li abundances for TO stars in NGC6397}
\label{ews}
\begin{center}
\begin{tabular}{lccccrrrrrrr}
\hline
\\
star \# & EW &  $\sigma_{MC}$ & EW$_{g}$ & $\sigma_{C}$ &S/N& A(Li) & $\sigma_{Li}$ & $\rm T_V$ & $\Delta T_V$& A(Li) & $\sigma_{Li}$\\
        & pm & pm             & pm       & pm           &   & T=6476K              &            & T=$\rm T_V$ & & \\
(1) & (2) & (3) & (4) & (5) & (6) & (7) & (8) & (9) & (10) & (11) &  (12) \\
\\
\hline
\\
\multispan{11}{\hfill  measures from our  VLT-UVES data \hfill}\\
\\
\hline
\\
1543 & 2.63   &  0.21 & 2.60  & 0.21 & 85  & 2.41 & 0.07 & 6401 & 30 & 2.36 & 0.05  \\
1622 & 2.04   &  0.16 & 2.01  & 0.18 & 105 & 2.29 & 0.07 & 6394 & 32 & 2.24 & 0.05  \\
1905 & 2.71   &  0.19 & 2.62  & 0.21 & 85  & 2.43 & 0.07 & 6456 &  9 & 2.41 & 0.04  \\
201432 & 2.06 &  0.17 & 2.11  & 0.21 & 131 & 2.29 & 0.07 & 6427 & 23 & 2.26 & 0.05  \\
202765 & 2.65 &  0.27 & 2.77  & 0.30 & 92  & 2.42 & 0.08 & 6394 & 32 & 2.36 & 0.06  \\
\\
\hline
\\
\multispan{11}{\hfill measures from archive VLT-UVES data \hfill}\\
\\
\hline
\\
A228  & 2.66  &  0.19 & 2.61  & 0.19 & 95  & 2.42 & 0.07 & 6274 & 58 & 2.28 & 0.06 \\
A447  & 2.53  &  0.21 & 2.66  & 0.25 & 73  & 2.39 & 0.07 & 6374 & 37 & 2.33 & 0.05 \\
A575  & 2.54  &  0.21 & 2.52  & 0.27 & 69  & 2.40 & 0.07 & 6362 & 40 & 2.32 & 0.05 \\
A721  & 2.13  &  0.24 & 2.30  & 0.36 & 51  & 2.31 & 0.08 & 6386 & 34 & 2.25 & 0.06 \\
A1406 & 2.66  &  0.34 & 2.84  & 0.42 & 43  & 2.42 & 0.09 & 6345 & 44 & 2.37 & 0.07 \\
A2084 & 2.89  &  0.23 & 2.88  & 0.38 & 48  & 2.46 & 0.07 & 6383 & 35 & 2.38 & 0.05 \\
A2111 & 3.35  &  0.16 & 3.29  & 0.21 & 86  & 2.53 & 0.06 & 6207 & 69 & 2.33 & 0.06\\ 
\\
\hline
\\
\multispan{11}{\hfill measures of \citet{pm96} from NTT-EMMI data \hfill}\\
\\
\hline
\\
A611 & 2.60  &  0.80  &       &      &    &       &      & 6371 & 29 & 2.34 & 0.18 \\
C4602 & 3.40 &  0.67  &       &      &    &       &      & 6306 & 39 & 2.42 & 0.11 \\
A853 &  4.30 &  0.50  &       &      &    &       &      & 6206 & 70 & 2.48 & 0.08 \\

\hline
\end{tabular}
\end{center}
\end{table*}

\begin{figure}[t!]
\resizebox{\hsize}{!}{\includegraphics[clip=true]{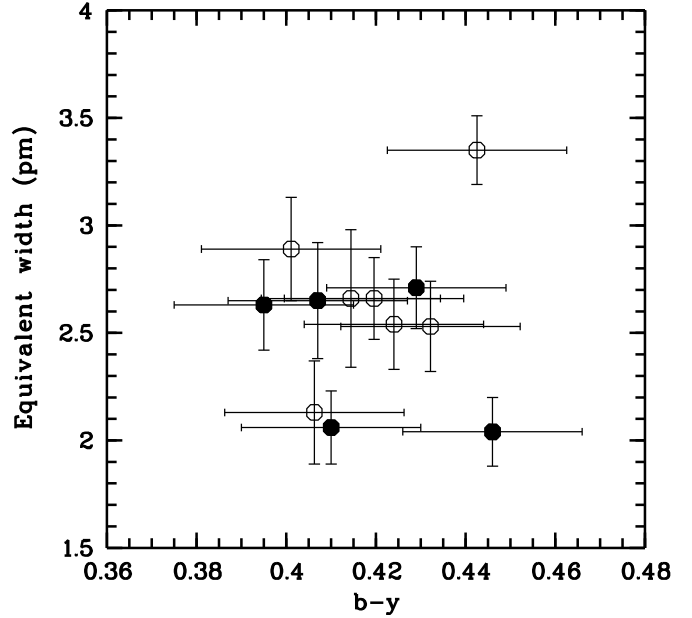}}
\caption{EW versus b-y. Filled symbols:
stars observed by us; 
open symbols: stars observed by \citet{thev} }
\label{by_ew}
\end{figure}

For each star we had four independent
exposures which were independently
calibrated, rectified  and then coadded. 
The coadded spectrum was normalized around the Li doublet 
by fitting a spline through several 
continuum windows about 0.2 nm ~ wide.

For the stars of \citet{thev} the raw data and calibration
frames were
retrieved from the ESO-VLT archive and were reduced
with exactly the same prescriptions as our own data. 
With respect to our own data  we highlight
the following differences of the \citet{thev} data: 1) the
CCD binning which was $1\times 1$; 2) the slit width 
was 1\farcs{1}; 3) the central wavelength 
was 580 nm, which implies that the Li doublet falls on
the MIT CCD; 4) there was no dichroic and observations were
taken in the red arm only.
The \citet{thev} data has a slightly lower resolution
and S/N, although a better sampling of the resolution element.
The flux calibration was performed using
the standard star  LTT9239, observed on 05/06/2000 with the
same instrument setup as the stars of NGC 6397.
Like for our own data multiple spectra of the same star
were coadded and normalized with a spline.

\section{Equivalent widths measurements}

In order to measure accurately the equivalent width of the Li doublet
we used a least-squares minimization
to determine the best fitting synthetic profile.
Since the data has been rebinned to a constant wavelength step
the theorems concerning the statistical
properties of the $\chi^2$ do not apply. Nevertheless
$\chi^2$ fitting may be used and the error
conveniently  estimated  using Monte Carlo simulations
(Bonifacio, in preparation).

To perform the minimization we used {\tt MINUIT} \citep{cern} using as fitting
parameters  equivalent width, broadening (assuming a gaussian
shape, which is appropriate for the UVES instrumental profile), 
wavelength shift
and a multiplicative constant for continuum adjustment.
To better constrain the continuum we included also
two continuum stretches of 0.2 nm each to the blue and to the red
of the line. 
The S/N ratio was estimated from the root mean square deviation
in these continuum windows and is given in 
column (6) of Table \ref{ews} . 
The fitted profiles are shown in Figures \ref{spectra} ,\ref{spectra4} and \ref{spectra3}
together with two bracketing synthetic profiles.
The equivalent widths of the fitted profiles 
in units of pm ($10^{-12}$m)
are given in column (2)
of
Table \ref{ews}. 
It is customary to use the Cayrel formula \citep{cayrel88} to estimate
the errors in equivalent widths. This formula assumes a gaussian line
profile and depends on the full width at half-maximun of the line, on the 
pixel size (in wavelength units) and on the S/N ratio.
Since the Li doublet profile is 
non-gaussian we performed a Monte Carlo simulation to estimate the 
errors on equivalent widths. For each star we took its best fitting
synthetic profile, added Poisson noise to obtain  the observed
S/N ratio
and performed the fitting several hundreds times.
The standard deviation of the fitted equivalent widths was then taken
as the 1 $\sigma$ error on the measured equivalent width.
The derived values are given in column (3) of Table \ref{ews}, 
the estimates from the Cayrel formula are given in column (5). 
This shows that the Cayrel
formula does provide a reasonable estimate of the error. 
We adopt the results of the Monte-Carlo simulations
as error estimates.
In column (4) of Table \ref{ews} 
we also provide the equivalent widths obtained by fitting
a gaussian to the data (we used the {\tt iraf} task {\tt splot}).
One may easily see that the two measures are consistent within 
errors. In fact at this resolution and S/N
approximating the doublet with a single gaussian is
acceptable.

\begin{table*}
\caption{Equivalent widths for field stars}
\label{field}
\begin{center}
\begin{tabular}{lrrrrrrrrrrr}
\hline
\\
star  & EW & $\sigma_{EW}$ &  [Fe/H] & log g & $T_{H\alpha}$ & $T_{F94}$& 
$T_{IRFM} $& A(Li) & $\sigma_{Li} $ & A(Li)$ $ & A(Li) \\
      & pm & pm & dex & cgs & K & K & K &  & & $Tc$ & ${TcNLTE}$ \\
(1) & (2) & (3) & (4) & (5) & (6) & (7) & (8) & (9) & (10) & (11) & (12) \\
\\
\hline
\\
HD 108177 & 3.37 & 0.09 &  -1.70 & 4.44  & 6027 & 6090 & 6067 & 2.25 & 0.11 & 2.28 & 2.29 \\
HD 116064 & 3.10 & 0.09 &  -1.87 & 4.35  & 6192 & 5822 & 5923 & 2.30 & 0.12 & 2.31 & 2.32 \\
HD 140283 & 4.95 & 0.06 &  -2.46 & 3.67  & 5898 & 5814 & 5691 & 2.31 & 0.12 & 2.36 & 2.38 \\
HD 166913 & 3.66 & 0.13 &  -1.59 & 4.08  & 5921 & 5955 & 6020 & 2.21 & 0.11 & 2.25 & 2.26 \\
HD 181743 & 3.78 & 0.09 &  -1.81 & 4.42  & 6130 &      & 5927 & 2.36 & 0.12 & 2.38 & 2.39 \\
\hline
\end{tabular}
\end{center}
\end{table*}

\section{Li Abundances}

\begin{figure*}[t!]
\resizebox{\hsize}{!}{\includegraphics[clip=true]{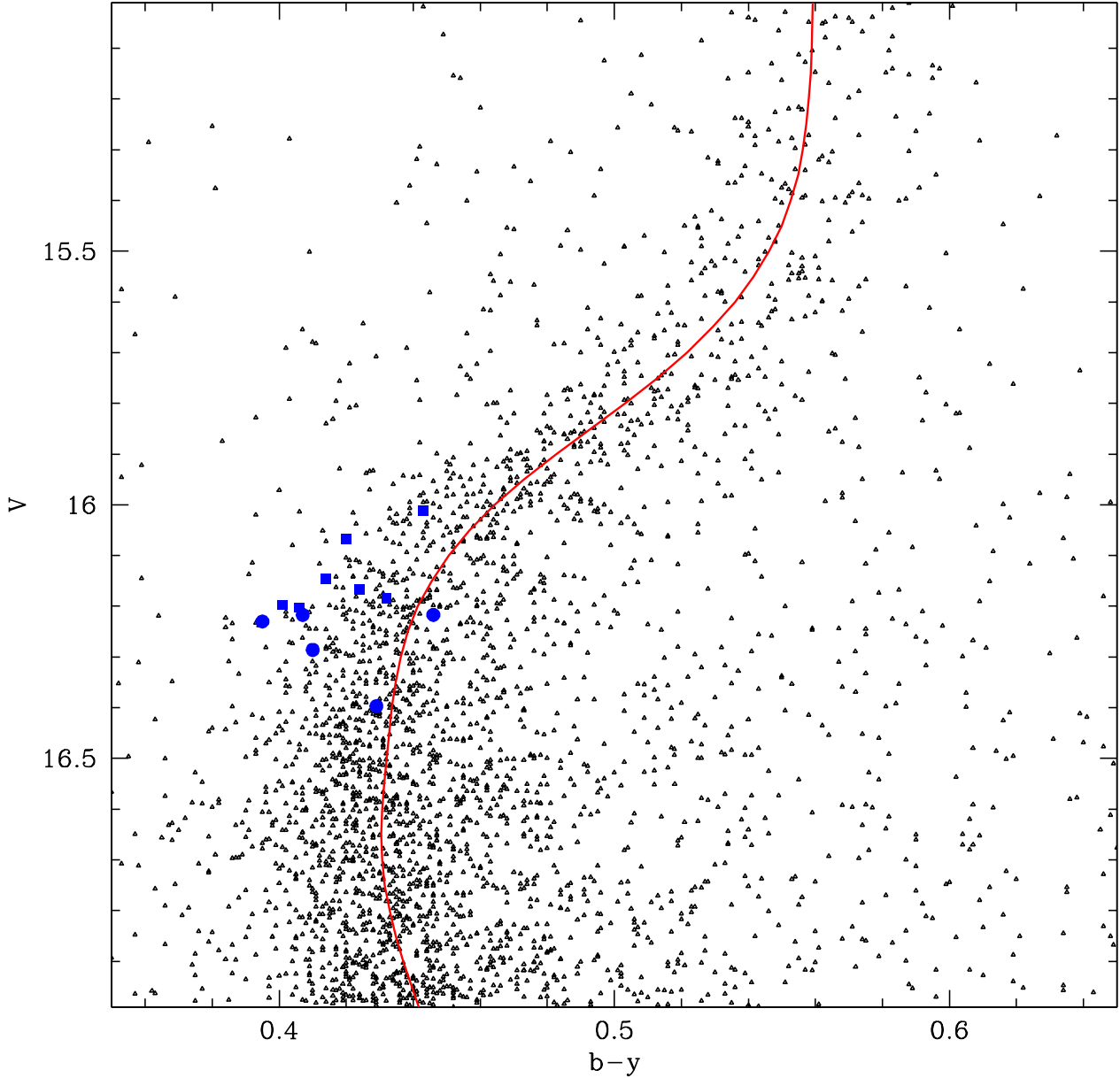}}
\caption{Color-magnitude diagram of NGC 6397, where
our own observations have been calibrated
using 3500 stars in common with \citet{twarog} .
Filled circles are the stars for which we
obtained UVES spectra and filled squares 
are those observed by \cite{thev} .
}
\label{cmd}
\end{figure*}

To determine the Li abundances 
we fitted the equivalent widths measured as described in the
previous section to those computed with the SYNTHE code
\citep{k93}, as described in \citet{bm97}.
The atmospheric parameters are the same adopted in \citet{g2001},
namely \teff = 6476 K ; log g = 4.10 and $\xi$ = 1.32 \kms and a
metallicity of $-2.0$ .
An LTE  model atmosphere with these parameters was computed with
the ATLAS 9 code \citep{k93} using $\alpha$ enhanced opacity
\footnote{We note that at this low metallicity the differences between
models with and without $\alpha- $enhancement are negligible
as shown e.g., by \citet{grat00}} 
distribution functions  with a microturbulent velocity of 1 \kms ~
and with the overshooting option switched off.
The results are given in column (7) of Table \ref{ews} .
In column (8) of the Table we 
provide also the errors in Li abundance obtained by summing
quadratically the errors arising from equivalent widths and from
effective temperatures; as error in \teff ~ we assumed 90 K, as given
in \citet{g2001}.
For this effective temperature the correction
for Li depletion predicted by the standard isochrones
of \citet{Deli} is zero.
The correction  for NLTE effects computed
by \citet{Carl}  is  +0.01 dex.

Although the assumption that all the stars share the same temperature 
is acceptable to determine the metallicity and the general
abundance pattern of the stars, it may prove to be somewhat
crude to discuss the dispersion of Li abundances.
In \citet{g2001} we derived the effective temperature
by fitting the mean  H$\alpha$ profile, obtained by
averaging all the spectra of the different stars, on
the assumption that they share the same temperature.
To derive temperatures for each star 
one could fit the individual H$\alpha$ profiles,
however, the relatively poor S/N 
of the spectra would not allow to derive 
temperatures with an accuracy better than 150 K.
Another possibility is to use $(b-y)_0$, but
this is even worse: a difference
of  $(b-y)_0$ of 0.015 mag translates into
a temperature difference of about 130 K.
The photometric error alone is of 
this order of magnitude. 
Figure
\ref{by_ew}  shows quite clearly that
there is no correlation between 
$(b-y)$ and Li equivalent width,
which should be the case if 
$(b-y)$ is tightly correlated with temperature
and the stars have more or less the same Li
abundance.
Instead our extensive photometry allows us to
derive quite accurately the mean locus 
of the cluster isochrone. What is shown in
Fig \ref{cmd} is  an average with outliers eliminated
using an iterative clipping procedure.

It is clear that all the stars observed are slightly
brighter than the turn-off where
a tight one to one relation holds
between V an $b-y$.
Adopting the average reddening
of $E(b-y)=0.137$, which we
derived comparing colours and temperatures of
cluster and field stars \citep{g2001}, we may thus map V onto
$(b-y)_0$ and hence temperature 
through comparison with the theoretical 
colors given by \citet{k93}.
These temperatures were corrected to be on the same
scale as those from H$\alpha$, exactly as done for
the field stars in \citet{g2001}. 
We recall that this temperature scale agrees
quite well with the IRFM temperature scale \citep{Alo96}.
We have applied this calibration to four field stars
with neglegible reddening, colours similar to our
TO stars and metallicities in the range $-2.2 \le \rm [Fe/H] -1.8$,
and found a mean difference of 7 K between
these temperatures and IRFM temperatures.
With this calibration an error of 0.05 mag in V
translates into an error in the range 20 -- 70 K.
These temperatures are given in 
column (9) of Table \ref{ews} 	and the errors
deriving from an error 0.05 mag are given in column (10).
This estimate is somewhat conservative
for the photometric error alone,
however if there is any variation
of reddening, even at the level 
of 0.01 mag in $E(B-V)$, this translates
into differences of 0.03 mag in V.
These temperature errors should be interpreted
as "internal errors", thus appropriate to assess
the issue of dispersion in Li abundances among the 
cluster stars. The "external error" which takes
into account the systematic error of our
adopted temperature calibration 
is difficult to assess, since it may depend on many
different sources, given that the calibration
involves several different steps, each with its
systematic error. 
However since we have shown we are on
the same scale of \citet{Alo96} we may assume
the error of this scale as systematic error.
This amounts to 80 K in this temperature range. 

The Li abundances derived assuming these temperatures
are given in column (11) of table \ref{ews}, and their
associate errors in column (12). 
For these temperatures the correction for Li  depletion 
predicted by the isochrones of \citet{Deli}
is zero for all but the two coolest stars, for which it is
of the order of 0.01 dex.
The NLTE corrections are also small and all of the order of
+0.01 dex for all stars.

\section{Discussion}

\subsection{Observed dispersion of Li abundances}

If we adopt the same temperature for all the stars
of NGC 6397
the mean Li abundance is
A(Li) = 2.39  \footnote{on the usual scale A(Li) = log [N(Li)/N(H)]+12},
with a standard deviation of 0.07 dex, or A(Li) = 2.40, once
NLTE effects are accounted for according
to \citet{Carl}. 
The question is whether 
such a scatter is consistent with the observational errors or if it
is larger.
The mean error is 0.074 dex, which intuitively
suggests that there is no extra scatter. 
A Kolmogorov-Smirnov test
of the null hypothesis that the data do {\em not} come 
from a normal distribution with mean 2.39 and standard deviation
0.074 gives a probability of $\sim 0.32$, thus rejecting the hypothesis.
One could argue that the Kolmogorov-Smirnov cannot
be applied here since the measures have different
errors and therefore cannot have been drawn from the 
same parent distribution.
We therefore test the hypothesis that the observed dispersion
is simply due to the observational errors by a straightforward
Monte Carlo simulation. We generate 1000 sets of 12 gaussian
random numbers with mean 2.39 and standard deviation equal
to the error estimate of each data point. 
For each realization of 12 ``observations'' we compute the
standard deviation $s$.
The mean value of $s$ over the sample is 0.073 with
a standard deviation of 0.015.
Hence the data is consistent with the notion that the stars
share the {\em same} Li abundance, the dispersion being
entirely due to observational error. 
The error on Li abundances is, in fact, largely dominated by the error
in \teff .
A spread in effective
temperature of 120 K, neglecting any error in the measurement of equivalent 
widths, would provide a spread in Li abundances of 0.08 dex, slightly
larger than the observed spread. This suggests that the error on the
effective temperature is in fact  10 or 20 K {\em larger} than estimated
in \citet{g2001}, which is not unreasonable.
It is therefore likely
that the stars do not share precisely 
the same effective temperature, as assumed
in \citet{g2001}.

\begin{figure}[t!]
\resizebox{\hsize}{!}{\includegraphics[clip=true]{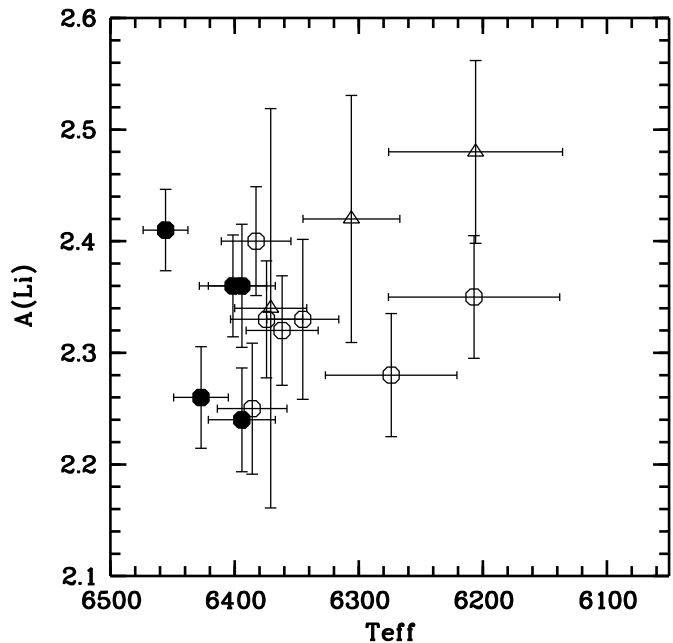}}
\caption{Li abundances 
as a function of effective temperatures
derived from the location of stars along the cluster mean locus.
The four stars observed by us are shown as filled circles,
those observed by  \citet{thev} as open symbols, 
while the three stars observed by \citet{pm96}
are shown as open triangles.
}
\label{li_teff}
\end{figure}

Let us therefore consider the sample with Li abundances derived
assuming the temperatures from  the $V - T_{eff}$ calibration.
It is apparent from the plot in Fig. \ref{li_teff}
that there is very little scatter. In fact also the three
stars observed with NTT by \citet{pm96} are totally
consistent with the present data set, once we adopt the same
temperature scale.
The mean  Li abundance is 2.32 with a standard deviation
of 0.056 dex. To verify whether this is compatible with
the estimated errors we resort again to a Monte Carlo
simulation. The mean standard deviation of 1000 samples is
0.051 with a standard deviation of 0.011.
Thus also with this temperature scale, in spite of the
very small errors adopted for the temperatures there
is no evidence of any intrinsic scatter.
In fact we may use these results to place an upper limit
on the maximum intrinsic scatter allowed.
In order to be not detected at 1$\sigma$ the intrinsic
scatter should have been such that, summed quadratically to the
mean scatter provided by observational errors (estimated
as 0.051 from the Monte Carlo simulation)
it does not exceed this mean by more than 1$\sigma$ .
Let $\delta \rm (Li)$ be the intrinsic scatter
produced by the Li depletion mechanism,
one has $\delta \rm (Li)< \sqrt{0.062^2 - 0.051^2} \approx 0.035$.
 
\subsection{Comparison with previous results}

\citet{pin99}
reported unpublished results by
Thorburn and collaborators
which showed  a significant 
dispersion of 0.3 -- 0.5 dex in Li
abundances in this cluster, based on spectra of about 20 stars, 
but
since these results have not been published in any 
detailed form we cannot make any hypothesis on the reason
for the discrepancy with the present results.

The difference between the
Li abundances given here and those
of \citet{thev} lies only in the different
effective temperatures  adopted.
Although \citet{thev} do not give 
the Li equivalent widths 
we  used their Li abundances 
and their effective temperatures to compute the
equivalent widths of the Li doublet for their stars,
and these were always within 0.2 pm of our measured
equivalent widths, as expected.
If we compare the colours for the two
sets of stars, by estimating for the Thevenin et al stars
$(b-y)\approx 0.74(B-V)$, we find that, although the
Thevenin et al stars are about 0.1--0.2 mag brighter,
thus  slightly more evolved
and, on average,   cooler,
they span the same colour range as our stars.
In Fig. \ref{by_ew} we show equivalent widths for our stars
(filled symbols) and for those of the \citet{thev} stars 
(open symbols)
as a function of $(b-y)$. It is clear that the two sets
of stars occupy the same area in this plane.

From Fig. \ref{li_teff} it is clear that our measures
are in good agreement with those of
\citet{pm96} and  the
difference in Li abundances is due to the
different temperature scales adopted.

\subsection{Comparison with field stars}

Several halo field stars have been observed at twilight in the course 
of the program and their atmospheric parameters, determined
as done for cluster stars are reported in \citet{g2001}.
The effective temperatures are different from those
of \citet{g2001} due to the better
straightening of the spectra achieved here thanks to the 
flux calibration.
To compare the Li content of NGC 6397 with that of
field stars we select only those with metallicity below
[Fe/H]$=-1.5$, which is roughly the value where 
Li abundances begin to rise off the {\em Spite plateau}, and
effective temperature higher than 5700, which is roughly
the value below which Li depletion due to convection
becomes non-negligible.
In Table \ref{field} we give the equivalent widths of
the Li doublet as measured from our spectra, the effective
temperatures derived from the H$\alpha$ profiles and
for reference, the effective temperatures of \citet{Fuhr94},
also based on Balmer lines, and of \citet{Alo96}, based
on the Infrared Flux Method (IRFM). 
The observed field stars are somewhat cooler than our cluster
stars. As noted by \citet{bm97} at temperatures below 6300 K,
even the {\em standard} models of stellar evolution
predict a slight Li depletion \citep{Deli}.
Thus to make a meaningful comparison we corrected the Li
abundances for the depletion predicted by the isochrones
of \citet{Deli} and these corrected values
are given in column (11) of Table \ref{field}: the corrections 
are small for all the stars. 
Also the correction for NLTE effects, albeit small, 
is temperature dependent \citep{Carl} we thus corrected also
for this effect and the corrected values are given
in column (12) of Table \ref{field}. 
As described above we
applied the same corrections also to the stars of
NGC 6397.

The star  HD 116064  has been noted to be a double-lined 
spectroscopic binary by \citet{smith} and our spectrum does indeed
confirm the duplicity.  The veiling is
unknown,  however the star's Li abundance
is comparable with that of the other stars, suggesting that at
these wavelengths it is neglegible. 
The mean A(Li) for the five field stars
is 2.29 with a standard deviation of 0.058,
in excellent agreement with 
the mean A(Li) $= 2.32$ for the 12   TO stars of NGC 6397.
If we consider
the depletion corrected values, the mean Li abundance is
2.31 for the field stars and 2.32 for the NGC 6397 sample. 
When considering also the NLTE corrections it becomes 2.33 for the 
field stars and 2.34 for the cluster stars.
We believe there is no need for any sophisticated statistical
analysis to conclude that the Li abundance
in NGC 6397 is the same as that of the field stars, within the stated
errors,
supporting the primordial origin of Li in these stars.

Preliminary results of
the analysis of a {\em high quality } sample of field
stars with IRFM temperatures  have been reported
by \citet{b01} and the current best estimate of
the plateau Li abundance is $2.317 \pm  (0.014)_{1\sigma}
\pm (0.05)_{sys}$. This value  happens to be
in perfect agreement with the values found here both
for field stars and for NGC 6397 TO stars.

\subsection{Evolutionary models and Li depletion}

There is a long-standing debate concerning Li depletion
on the {\em Spite plateau}.
If Li depletion is possible, the pristine 
Li abundance is higher than the observed.
As noted above, {\em standard} models of stellar
evolution cannot account for any significant 
depletion in dwarf stars.
Early attempts to compute more sophisticated
models including diffusion \citep{mich84} or
rotation-induced turbulence \citep{vac88}
predicted large depletions\footnote{For the
purpose of the present discussion we consider
Li to be depleted, even if it is not destroyed,
but simply removed from the observable layers
of the stars. What is relevant is that
the true Li abundance is higher than that deduced
from the analysis of the spectrum.}
of up to a factor
of 10. These models predicted a considerable scatter
in the observed Li abundances 
and also other features, 
such as a slope 
of the plateau with effective temperature and 
metallicity, or an enhanced Li abundance among post-TO 
stars. All of these
effects  have been ruled out by the observations. 

The situation 
on the theoretical side has  
dramatically changed in the last year:
new sets of models seem to be able to produce 
a plateau with very 
little scatter.
The rotationally mixed models
of \citet{pin01}, the diffusive models
of \citet{sal01} and the models
of \citet{theado}, which include rotational-mixing,
diffusion and compositions gradients, are all capable
of reproducing a plateau, with a very small intrinsic
dispersion. It must be however noted, that large
depletion factors are not any longer found,
and  the above papers predict primordial Li abundances
in good agreement with the best current estimates
of the baryonic density of the universe \footnote{
 \citet{pin01} derive A(Li) = 2.4, \citet{sal01} and \citet{theado}
derive A(Li) = 2.5}.
The intrinsic dispersions allowed for
by the various depletion mechanisms, are somewhat larger
than the maximum value of 0.035 dex derived above; 
it is unlikely, however, that this value can rule out
any of them.
A common feature of all models which predict Li depletion
is the presence of a small number of ``outliers'',
i.e. stars with a Li abundance more than $3\sigma$
below the mean and such objects  are indeed known to exist
among field stars. It may therefore of some
significance that out of 15 TO stars observed 
in NGC 6397 none appears to be significantly
Li depleted.
We mantain that  there is no compelling evidence
in favour of any of the proposed depletion
mechanisms, although none can be ruled out.
Economy of hypothesis suggests that 
{\em standard} models with no Li depletion
should be preferred.
In spite of this  there are strong arguments
in favour of the presence of both diffusion and
rotational mixing,
although, admittely, they come 
mostly from the study of the Pop I. 

Helioseismic studies of the Sun 
imply that atomic diffusion is important \citep{gue96}. Theoretical
investigations show that it should be important also
in metal-poor stars \citep{mich84,cast97,sal01}.
\citet{leb99} have shown that 
the H-R diagram 
of field stars with accurate parallaxes, in the metallicity range
$-1.0\le+0.3$,
is best fitted when models which include diffusion
are used. 
All diffusive models predict that surface lithium 
sinks below the bottom of 
the thin surface main sequence convection zone, thus
becoming unobservable, i.e.  depleted, in our parlance.
Two  features are predicted by diffusive models: excess
dispersion and a downturn at the hot end of the plateau;
both are in disagreement with the observations.
To alleviate the problem it has
been suggested that a modest mass-loss \citep{vc95,swen95}
may undo the sedimentation predicted by diffusion.
These mass loss models, however, also predict
a down-turn of
Li abundances at the hot edge of the {\em Spite plateau}
\citep{vc95} and a constant Li abundance among
cooler sub-giants \citep{swen95}, but
none of these features is
observed.
Furthermore there is no observational evidence that such
a mass-loss actually occurs.
Recently \citet{vac99} has proposed a model in which 
meridional circulation, in presence of a composition gradient, 
created by diffusion,
could lead to a
quasi-equilibrium stage in which diffusion and
circulation are both inhibited.
The detailed results for these model and the impact
on the {\em Spite plateau} have been discussed
by \citet{theado}.  
The predicted dispersion in the plateau is very small,
however there persists a slight downturn at high 
temperatures and an increased dispersion.
Similar results are found by 
\citet{sal01},  using purely diffusive
models (no rotational mixing and no
composition gradients): a  very
uniform  plateau, with a slight 
downturn, and excess dispersion at the hottest 
edge, above 6200 K (see their figures 7 and 8).
This is exactly the region covered by our observations
and no downturn or extra dispersion are observed.
Admittedly our sample, of 12 stars only, is very small,
and \citet{sal01} predict, based on their Monte Carlo simulations,
that at least 40 -- 50 stars at the hot end of the
{\em Spite plateau} should be observed
in order to detect the downturn. 
Such a number of stars shall be easily available
in the near future, with the FLAMES instrument,
but for the time being we conclude that 
our results for NGC 6397 do not support
neither the models of \citet{theado} nor those
of \citet{sal01}.
The Li abundance in NGC 6397
is constant 
and is the same
as that of  field stars of much lower \teff.
Also the iron abundance of the
TO stars  \citep{g2001,thev} is equal to  that 
of sub-giant and giant stars \citep{g2001,c00}. 
Instead diffusive models predict the
photospheric metallicity of a star  to constantly
decrease, due to diffusion, during its main sequence
lifetime, reaching a minimum around the turn-off
and then increasing back again to the original
metallicity along the red giant branch \citep{cast97}.
The observational evidence presented here and in
\citet{g2001} is against any significant 
sedimentation due to diffusion
in the stars of NGC 6397, both for Li and Fe.

The strongest evidences 
of the inadequcy of
the standard models probably comes from 
the study of lithium in open clusters; among them, the presence of the 
Li dip first discovered in the Hyades 
\citep{1987ApJ...313..389B},
the observed scatter in Li abundance among the solar type
stars in the field and 
M67 \citep{1994A&A...287..191P,1997A&A...325..535P},
require that additional mixing mechanisms, besides  convection,
are at work.   
Rotational mixing models have the ingredients to reproduce 
several of the observed features, and, although they  fail in the 
quantitative reproduction of the observations (see e.g. 
\citet{2000A&A...356L..25R,Randich2002}),
they are  promising,
expecially when considering stars with high masses and large rotational
velocities such those around the Li dip. 
Open cluster data has been interpreted with
models with different physical ingredients:
\citet{pasquini01} used
the \citet{1999A&A...351..635C}
models, which include both rotational mixing
and diffusion,
for the interpretation of the Li observations of 
the intermediate age cluster NGC 3680; 
\citet{2002ApJ...565..587B}
used the 
\citet{1997ApJ...488..836D}
models, which include only rotational mixing, 
for the interpretation of the 
Hyades Li and Be data. 
Therefore there is no
``non standard'' set of models which is prefered
for the explanation of Li observations  in Pop I stars.
We point out that 
the Pop II stars are of much lower metallicity and
mass. We therefore 
do not see any compelling argument for the use
of ``non standard'' models, in the case of Pop II stars. 
The only observable prediction we could verify  is that they predict the 
presence of some `'outlier'' on a large sample; in a sample of 15 
NGC 6397 turnoff stars we do not find any outlier, and this 
could be therefore taken as a mild indication against 
extra mixing predicted by ``non standard'' models. 
From an observational point of view we 
wish to add that  \citet{sara} showed that  the
projected rotational velocities of our stars 
are less than 1.7 \kms, which
provides an
additional constraint, to  be incorporated
in the comparison of theoretical and observational data.

\subsection{Normal Li and enhanced N: a  paradox ?}

The subgiant stars of the cluster show very strong CN
bands. The full analysis of this data shall be
presented elsewhere, 
however preliminary results suggest a
conspicuous N enhancement of the order
of 1 dex or more. This  suggests
that the material has been processed through 
the CNO cycle. Since our current understanding 
of subgiant stars precludes any mixing to occur
in the star itself one is forced to
conclude that the star was either
formed from, or polluted by, such processed material.
One should therefore expect that the TO
stars share the same characteristics, although
the CN bands are too weak to be detected and
no direct information on N abundance is available.
Here lies the paradox: if the material
of these stars has been processed at the high
temperatures where the CNO cycle is
operating, Li should have been destructed.

Such a paradox has been known to exist
for many years for the N-rich metal-poor
dwarfs, like HD 166913, studied also 
in this paper, which are Li - normal
\citep{ss86}. 
In two of these stars \citet{bs94}
have found the neutron capture elements 
to be moderately enhanced and concluded
that the most likely source of N and
neutron capture elements was material
provided by a thermally pulsing AGB star.
In the
stars of NGC 6397, instead, [Sr/Fe]$\sim -0.1$,  
[Ba/Fe]$\sim -0.2$
and [Eu/Fe]   $\sim 0.4$,
as shall be discussed elsewhere.
\citet{bs94} also  argued that 
the N enhancement in halo dwarfs is
unrelated to the phenomenon 
observed in N-, Na-, Al-rich, O-poor
globular cluster giants, on the grounds that
oxygen is not depleted in N-rich dwarfs.
The  abundance ratio of oxygen and magnesium to iron 
in this cluster
is rather low $\sim +0.2$.
Both oxygen and magnesium 
deficiencies could 
be explained by presence of material processed through
complete CNO cycle.
We add that Al enrichment is very moderate
or absent in N-rich dwarfs, at the level
of [Al/Fe]$\sim 0.2$ at most \citep{bs94,f86}.
Our TO stars in NGC 6397 show a similar
low Al enhancement. 

Our current understanding 
of N production require that it
is synthetized in intermediate mass
stars through CNO cyclying \citep{2000ApJ...541..660H}.
However the simple picture by which Li is totally
destroyed under these condition is not confirmed
by recent computations.
\citet{Ventura} studied the evolution of N and Li in
metal-poor massive AGB stars,
CNO processing at the bottom of the
convective envelope
overproduces nitrogen by $\sim 1.5$ dex;
lithium is initialy produced by
the Cameron-Fowler mechanism 
\citep{cameron} and then
destroyed, 
but a small amount of matter very Lithium rich is 
recycled to the interstellar medium.
The net result
is a reduction  of 
 $\sim 0.6$ dex
for stars of metallicity 
comparable to that of NGC 6397.
Note that a mass loss rate 
four times larger than the one assumed
(which is within the 
uncertainty in the mass loss calibration) 
would leave the lithium essentially unaltered.
\citet{2001ApJ...554.1159C} have computed models
for the evolution of zero metallicity stars.
Such intermediate mass stars are capable
of producing primary N and the most
massive ones experience a normal AGB
evolution, including thermal pulses
and third dredge--up.
What is interesting for us here,
is that also in this case
there is Li production via the
Cameron-Fowler mechanism.
We conclude that there are strong theoretical indications
that at low metallicities N production is always
accompanied by some Li production, so that Li is never
totally depleted in the processed material.  
Mixing such material with 
material having the big bang
lithium abundance would dilute nitrogen 
leaving it overabundant, but would
not touch significantly lithium. 
Whether this
is indeed a viable explanation depends very much
on the precise degree of nitrogen enhancement.
This may somehow alleviate the paradox, however
some  fine tuning is still needed in order
to produce a Li abundance 
exactly at the level of the {\em Spite plateau}.
 
We are not here in a position to solve this
puzzle, however we conclude this section
by reasserting what we believe to be
a sound observational fact:
whichever the cause of these abundance
anomalies, it cannot have altered significantly 
the Li content of the cluster stars.

\section { Conclusions }

The equivalent widths of the Li doublet 
of the 12 turn-off stars of NGC 6397 observed 
with VLT+UVES are consistent
with the hypothesis that the stars all share the same Li  abundance,
the spread being entirely explained  by observational error.
The stars do not have exactly the same temperature
but span a range of almost 300 K.
We thus consider the chemical homogeneity of this cluster,
which we found for iron abundances in \citet{g2001}, to be
firmly established.
Why a globular cluster like NGC 6397 should show a constant Li abundance,
with essentially no intrinsic scatter, while M92, M71 and M13 show 
a scatter is puzzling. We note here however, that while M92 is a very 
metal--poor cluster, comparable to NGC 6397, both M71 and M13 are rather
metal-rich. In fact more metal rich than [Fe/H]$=-1.5$, which is roughly the
metal-rich edge of the {\em Spite plateau}.

The constancy of Li abundances within the cluster has two important
implications:
\begin{enumerate}
\item it poses a very stringent upper limit to the maximum intrinsic dispersion
      allowed for any 
      proposed Li depletion  mechanism,
      which must be less than 0.035 dex.
\item the good agreement between the mean Li abundance in the
      cluster and that obtained for the field stars suggests that this
      is in fact the primordial Li abundance.
\end{enumerate}

The mean value of  A(Li)
in the cluster, based on 12 stars, including
corrections for ``standard'' depletion and NLTE effects, 
is A(Li)$=2.34$
with a standard deviation of 0.056.
This is a statistical error to which we must add
the  systematic error corresponding to 80 K in the temperature
scale, which amounts to 0.06 dex in Li abundance.
Taking this 
value as the estimate of the primordial Li abundance, we obtain
two possible values for the baryon-to-photon ratio:
$n_b/n_\gamma =4.3 \times 10^{-10}$ which, using the relation
$\Omega_bh^2 = 3.66\times 10^{-3}\times \eta\times 10^{-10}$,  
implies $\Omega_bh^2 = 0.016\pm 0.04$, in good agreement
with recent results  from the  abundance of D
\citep{dodo,lev} and $\rm ^3He$ \citep{bania} and from the CMB fluctuation
spectrum \citep{debe};
or $n_b/n_\gamma = 1.5 \times 10^{-10}$ which implies
a low baryon density $\Omega_bh^2 = 0.005^{+0.0026}_{-0.0006}$.
In spite of the considerable error on the Li abundance,
the error on $\Omega_b$ is dominated by the error
on the theoretical predictions of nucleosynthesis.

The lithium content of NGC 6397 supports the primordial
origin of lithium  in metal-poor stars
and the standard big-bang model without need for any new physics.

\begin{acknowledgements}

Part of this work was done while PB was
at the Observatoire de Paris-Meudon as
a visitor.
R. Cayrel and M. Spite are warmly  thanked for many
helpful discussions. 
This research was done with support from the 
italian MURST/MIUR COFIN2000 grants for the
projects ``Osservabili stellari di interesse cosmologico''
(P.I. V. Castellani) and
``Spettroscopia e studio dei processi fisici del Sole e delle 
stelle di tipo solare'' (P.I. G. Peres)

\end{acknowledgements}

\bibliographystyle{aa}

\end{document}